\documentclass[twocolumn,showpacs,aps,prl]{revtex4-1}

\usepackage{mathrsfs}
\usepackage{amssymb, amsbsy, amsmath, latexsym, dsfont, array, layout, mathrsfs, color}

\usepackage{mathtools}

\usepackage{amsmath,amssymb}
\usepackage{bm}
\usepackage{graphicx}
\usepackage{braket,dsfont}
\usepackage{color}
\usepackage[10pt]{moresize}
\usepackage{mathrsfs}
\usepackage{multirow}
\usepackage{hyperref}
\usepackage{changes}
\usepackage{mathdots}
\let\oldcaption\caption
\renewcommand{\caption}{\sffamily \oldcaption}

\usepackage{amsmath,amsfonts,amssymb}
\usepackage{wrapfig}
\usepackage{graphicx}
\usepackage{bbm}
\usepackage[normalem]{ulem}
\usepackage{color}

\begin{document}

\title{Counting Classical Nodes in Quantum Networks}\date{\today}

\author{He Lu$^{1,2,3}$}
\thanks{These two authors contributed equally to this work}
\author{Chien-Ying Huang$^{4}$}
\thanks{These two authors contributed equally to this work}
\author{Zheng-Da Li$^{1,2}$}
\author{Xu-Fei Yin$^{1,2}$}
\author{Rui Zhang$^{1,2}$}
\author{Teh-Lu Liao$^{4}$}
\author{Yu-Ao Chen$^{1, 2}$}
\email{yuaochen@ustc.edu.cn}
\author{Che-Ming Li$^{4,5,6}$}
\email{cmli@mail.ncku.edu.tw}
\author{Jian-Wei Pan$^{1, 2}$}
\email{pan@ustc.edu.cn}

\affiliation{$^1$Shanghai Branch, National Laboratory for Physical Sciences at Microscale and Department of Modern Physics, University of Science and Technology of China, Shanghai 201315, China}
\affiliation{$^2$Synergetic Innovation Center of Quantum Information and Quantum Physics, University of Science and Technology of China, Hefei, Anhui 230026, China}
\affiliation{$^3$School of Physics, Shandong University, Jinan 250100, China}
\affiliation{$^4$Department of Engineering Science, National Cheng Kung University, Tainan 701, Taiwan}
\affiliation{$^5$Center for Quantum Frontiers of Research \& Technology, National Cheng Kung University, Tainan, 701, Taiwan}
\affiliation{$^6$Center for Quantum Technology, Hsinchu, 30013, Taiwan}

\begin{abstract}

Quantum networks illustrate the use of connected nodes of quantum systems as the backbone of distributed quantum information processing. When the network nodes are entangled in graph states, such a quantum platform is indispensable to almost all the existing distributed quantum tasks. Unfortunately, real networks unavoidably suffer from noise and technical restrictions, making nodes transit from quantum to classical at worst. Here, we introduce a figure of merit in terms of the number of classical nodes for quantum networks in arbitrary graph states. Such a network property is revealed by exploiting a novel Einstein-Podolsky-Rosen steerability. Experimentally, we demonstrate photonic quantum networks of $n_q$ quantum nodes and $n_c$ classical nodes with $n_q$ up to 6 and $n_c$ up to 18 using spontaneous parametric down-conversion entanglement sources. We show that the proposed method is faithful in quantifying the classical defects in prepared multiphoton quantum networks. Our results provide novel identification of generic quantum networks and nonclassical correlations in graph states.

\end{abstract}

\maketitle
\begin{figure*}[t!]
\centering
\includegraphics[width=1.8\columnwidth]{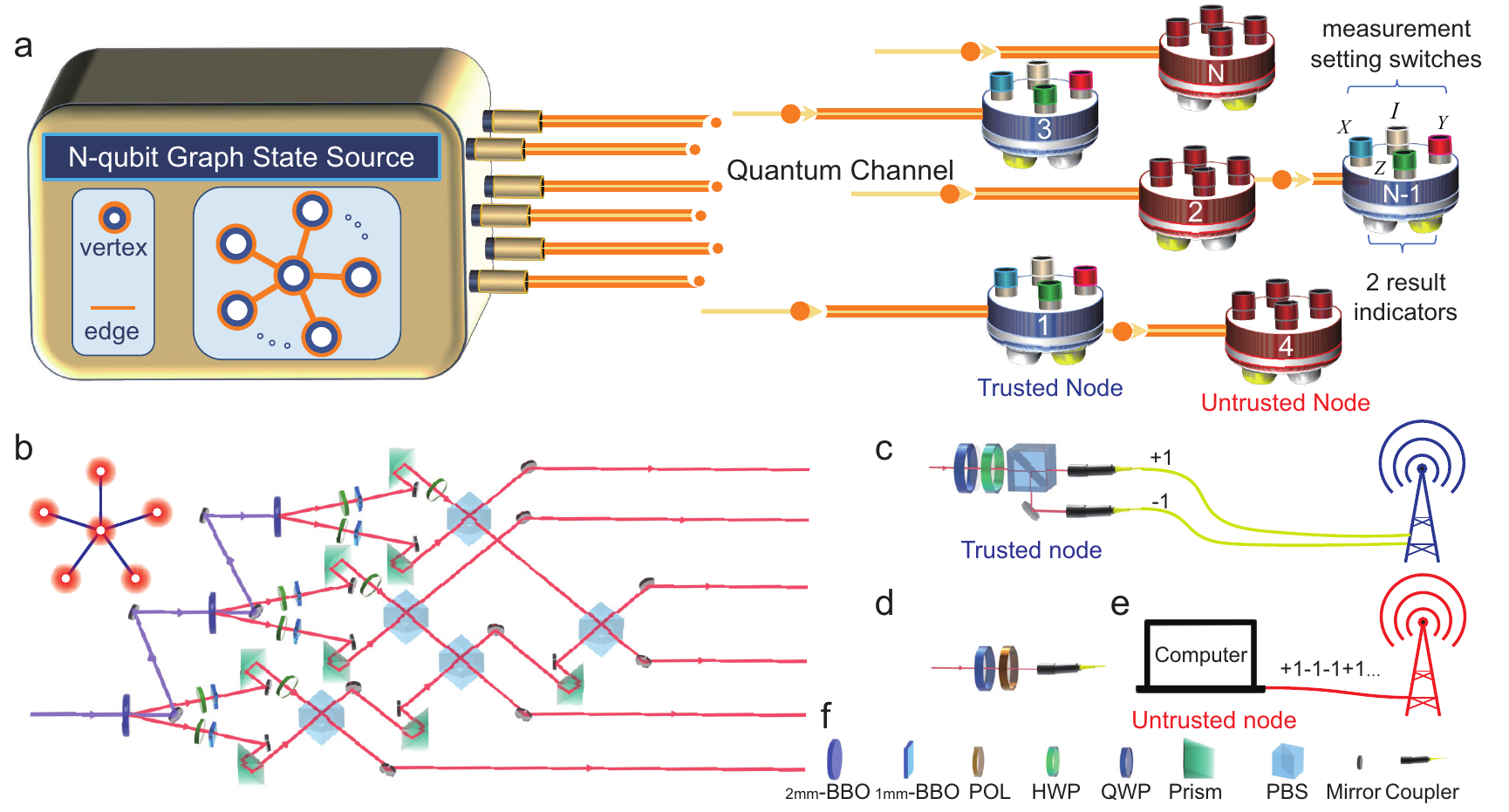}
\caption{
Schematic drawing of the quantum network in graph state and its experimental realization. (a) A quantum network ideally prepared in a graph state is depicted using the graph $G(V,E)$ \cite{Hein04,SM}. The graph $G$ consists of the vertex set $V$ and the set $E$ of edges each of which joins two vertices. The vertices and the edges physically represent the qubits and the interacting pairs of qubits respectively, and then constitute a state vector $\left|G\right\rangle$ of the network. A quantum network in graph state $\ket{G}$ is then distributed to distant nodes and verified by measurement apparatus. The measurement setting is chosen from set $\{I, X, Y, Z\}$, each of which has two outcomes $+1$ and $-1$. The blue (red) measurement apparatus represents trusted (untrusted) nodes in quantum network. It has been shown that arbitrary graph states among the network participants for distributed tasks can be established through a modular and plug-and-play architecture \cite{Pirker17}. (b) The experimental setup to generate a six-photon state in star graph, which is equivalent to $\ket{GHZ}_6$ via LOCC. (c) The experimental setup to measure network fidelity $F$. (d) The experimental setup to generate the state in the optimal ``cheating strategy", in which we project one photon on $\ket{\xi^{\prime}}$ according to the target state $\ket{G}$. (e) The untrusted node broadcasts results according to measurement setting of $F(6)$ \cite{SM}. (f) Symbols used in (b), (c) and (d): 2mm-long BBO crystal (2mm-BBO), 1mm-long BBO crystal (1-mm BBO), polarizer (POL), half-wave plate (HWP), quarter-wave plate (QWP), and polarization beam splitter (PBS).
}
\label{fig:setup}
\end{figure*}

Quantum mechanics enables nonclassical correlations to exist across the whole of a network via connecting individual quantum nodes, forming a joint quantum many-body system \cite{Ritter12}. Quantum networks \cite{Kimble08,Wehner18} have far greater capacity than the classical ones and serve as well-advanced transmitters of quantum information for all the distant network participants. Such utilities encourage important applications in distributed quantum information processing, from quantum secret sharing (QSS) \cite{Hillery99,Chen05,Markham08,Bell14,Lu16,Huang19} to distributed sensing \cite{Komar14,Proctor18}, and from distributed quantum computation \cite{Raussendorf01,Walther05,Broadbent09,Barz12} to quantum conference key agreement and distribution \cite{Chen07,Lo14,Epping17}. The physical realization of these distributed
quantum tasks requires suitable connectivities between nodes and network topologies to initialize the nodes in the multipartite entangled states, known as \emph{graph states} \cite{Hein04} [see Fig.~\ref{fig:setup}(a)].

To establish a quantum network in graph states with tailored topology, quantum information demands to be sent, received, stored, and exchanged between remote quantum nodes via photonic channels in general \cite{Kimble08,Wehner18,Ritter12,Northup14,Monroe14,Reiserer15,Sipahigil16,Kalb17,Pirker17,Humphreys18,Chou18,Jing19,Yu20}. Then, it is essential to characterize a created network before it
carries out a given distributed task, such as a QSS scheme. A conventional way to detect entanglement in the laboratory is an entanglement witness (EW), which employs deduction from the predictions of quantum theory \cite{Horodecki09,Guhne09,McCutcheon16}. However, inevitable imperfections of network nodes, such as the intrinsic fragility of quantum systems and errors present in actual implementations, can cause quantum nodes to become classical systems that obey the laws of classical physics, therefore leading to the failure of state preparation or decay of quantum networks \cite{Kimble08,Wehner18,Ritter12,Northup14,Monroe14,Reiserer15,Sipahigil16,Kalb17,Pirker17,Humphreys18,Chou18,Jing19,Yu20}. Moreover, when network participants only have limited knowledge about the node imperfections, the network nodes then become untrusted to the participants as untrusted nodes. Considering the existence of untrusted nodes in the created network, the EW is no longer reliable in verification of multipartite entanglement. This raises a natural question: How can a verifier, such as the dealer in QSS, objectively and reliably detect the presence of classical nodes in a given network for distributed tasks?

In this Letter, we address this issue by exploiting a novel Einstein-Podolsky-Rosen (EPR) steerability \cite{Wiseman07,Jones07, Li15}, which is capable of excluding the existence of classical nodes in quantum networks. More importantly, the EPR steerability presents more fine-grained information about the created network, i.e., the capability of counting the
number of classical nodes in the created network, which is not possible in other schemes \cite{Pappa12,McCutcheon16}.

Given an ideal $N$-node quantum network in arbitrary graph state $\ket{G}$, where each node contains a qubit, its general state decomposition can be explicitly expressed as \cite{Nielsen10,James01}
\begin{equation}
\left|G\right\rangle\!\!\left\langle G\right|=\sum_{\vec{m}}h_{\vec{m}}\bigotimes_{k=1}^{N}\hat{R}_{m_{k}},\label{qstm}
\end{equation}
where $h_{\vec{m}}$ are coefficients and $\hat{R}_{m_{k}}$ represents the $m_{k}$th observable of the $k$th node. We then introduce the network fidelity function for arbitrary target graph states $\left|G\right\rangle$ of $N$ nodes
\begin{equation}
F(N)=\sum_{\vec{m}}h_{\vec{m}}\left\langle R_{m_{1}}...R_{m_{N}}\right\rangle\label{fg}
\end{equation}
where $\vec{m}\equiv (m_{1},...,m_{N})$ and $R_{m_{k}}$ is the outcome of the $m_{k}$th measurement of $\hat{R}_{m_{k}}$ on the $k$th node. In our study, the measurements on each qubit are performed with the observables in Pauli matrices, $\{\hat{R}_{m_{k}}|m_{k}=0,1,2,3\}$, where $\hat{R}_{0}=I$, $\hat{R}_{1}=X$, $\hat{R}_{2}=Y$, and $\hat{R}_{3}=Z$. Note that the fidelity function~[Eq.~(\ref{fg})] is state dependent and we obtain $F=1$ for an ideal quantum network regardless of what fidelity function is chosen. We utilize the network fidelity function, which measures the closeness of created networks and target graph states, as the basis for counting classical nodes. This makes our framework capable of being used in a wide variety of circumstances and applications based on the fidelity measure.

When classical nodes exist in the created network, the network becomes a hybrid system consisting of $n_q$ quantum nodes and $n_c$ classical nodes, where $N=n_q+n_c$. The index set of the network nodes, $V$, can then be divided into the quantum-node subset, $V_{Q}$, and the classical-node subset, $V_{c}$, accordingly. An essential difference between quantum and classical nodes is that physical properties of quantum nodes might not have definite values. In contrast, variables in classical nodes are in existing states independent of observation, known as the assumption of \emph{realism} \cite{Mermin93,Guhne05,Brunner14}. In our framework, a node is defined as being \emph{classical} if, for any physical properties of interest, it is classical realistic, i.e., the state of each classical node can be specified by a pre-existing and fixed set of measurement outcomes \cite{Li15}. Note that with the increase of noises, the quantum nodes can eventually be described by the classical realistic theory. See Supplementary Material (SM) for detailed discussion of classical nodes \cite{SM}.

\begin{figure*}[t!]
\centering
\includegraphics[width=1.8\columnwidth]{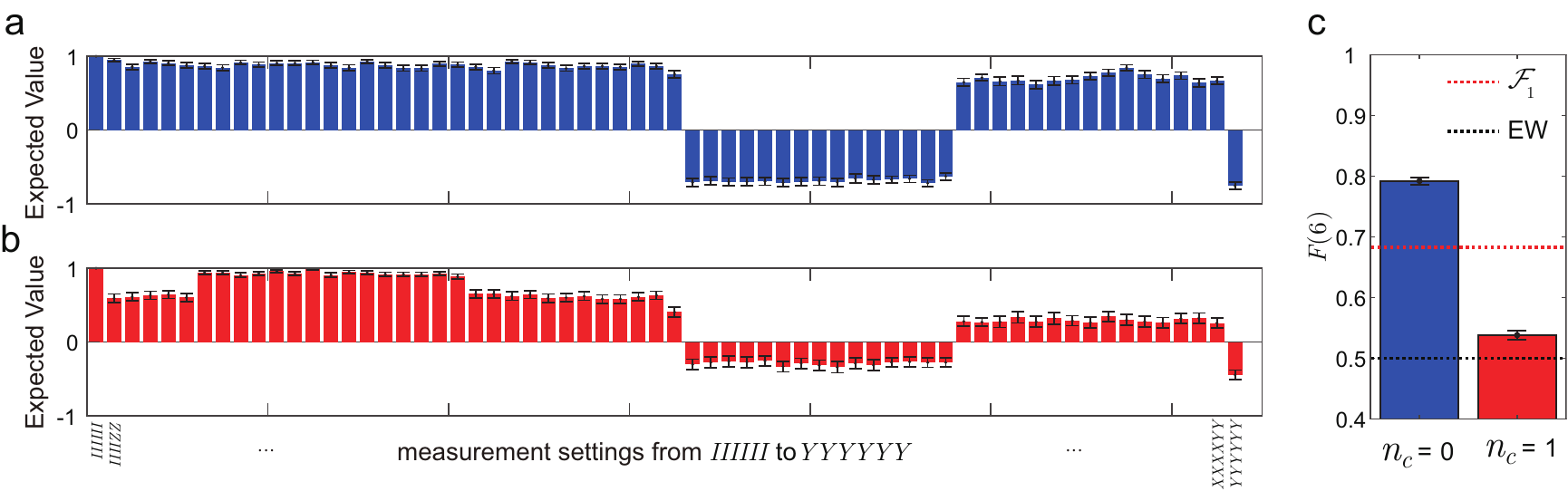}
\caption{Experimental results of network fidelity $F(6)$ of six-node network with classical node number $n_c=0$ and $n_c=1$ respectively. (a) The experimental results of network fidelity in network with $n_c=0$. (b) The experimental results of network fidelity measurement in network with $n_c=1$. (c) The calculated $F(6)$ from the results in (a) and (b). The black dash line is the threshold of EW, and the red dash line is the threshold fidelity $\mathcal F_1$. 
}
\label{fig:data1}
\end{figure*}

Before the state decays, the state vector of the target graph state can always be represented in the Schmidt form of rank $r$ \cite{Li10}
\begin{equation}
\left|G\right\rangle=\frac{1}{\sqrt{r}}\sum_{v=0}^{r-1}\left|v\right\rangle_{sQ}\left|v\right\rangle_{sc},\label{SD}
\end{equation}
for $r\geq2$, where $\{\left|v\right\rangle_{sQ}\}$ and $\{\left|v\right\rangle_{sc}\}$ are the Schmidt bases for the nodes in the vertex sets $V_{Q}$ and $V_{c}$, respectively. This representation shows us the following form under the state decomposition [Eq.~(\ref{qstm})] for the nodes in $V_{c}$
\begin{equation}
\left|G\right\rangle\!\!\left\langle G\right|=\frac{1}{r}\sum_{v,v',\vec{m}_{c}}h_{\vec{m}_{c}}^{vv'}\left|v\right\rangle_{\!sQsQ}\!\!\left\langle v'\right|\bigotimes_{k\in V_{c}}\hat{R}_{m_{k}}, \label{StDecom}
\end{equation}
where $\left|v\right\rangle_{\!scsc}\!\!\left\langle v'\right|=\sum_{\vec{m}_{c}}h_{\vec{m}_{c}}^{vv'}\bigotimes_{k\in V_{c}}\hat{R}_{m_{k}}$, $\vec{m}_{c}\equiv \{m_{k}|k\in V_{c}\}$, and $h_{\vec{m}_{c}}^{vv'}$ denote the decomposition coefficients. With the classical realistic theory for a complete description of the total state of the $n_c$ classical nodes in terms of the preexisting outcome sets: $\{\textbf{v}_{k}|k\in V_{c}\}$, the network fidelity function [Eq.~(\ref{fg})] can be rephrased as the following explicit form:
\begin{equation}
F=\frac{1}{r}\sum_{v,v',\vec{m}_{c}}h_{\vec{m}_{c}}^{vv'}\left\langle\left|v\right\rangle_{\!sQsQ}\!\!\left\langle v'\right|\right\rangle\left\langle\prod_{k\in V_{c}}R_{m_{k}}\right\rangle.\label{F}
\end{equation}

The maximum fidelities between target graph states $\ket{G}$ and $N$-node networks having $n_{c}$ classical nodes can then be described by the equation 
\begin{equation}
\mathcal{F}_{n_{c}}=\frac{1}{4}(1+2^{\frac{-n_{c}}{2}}\sqrt{4+2^{n_{c}}}), \label{fnc}
\end{equation}
where $1\leq n_{c}\leq N-1$, holding for arbitrary target graph states (See SM for the detailed derivation \cite{SM}). The threshold fidelities $\mathcal{F}_{n_{c}}$ strictly decrease with the number of classical nodes $n_{c}$. It turns out that there exists a one-to-one correspondence between the number of classical nodes and the relevant maximum fidelity values. For instance, $\mathcal{F}_{1}\simeq0.6830$, $\mathcal{F}_{2}\simeq0.6036$, and $\lim_{n_{c}\rightarrow\infty}\mathcal{F}_{n_{c}}\simeq0.5000$. The hybrids of quantum and classical nodes are then comparable in fidelity to the netwroks composed entirely of quantum nodes with $F\leq\mathcal{F}_{1}$. This implies that the collection $\{\mathcal{F}_{n_{c}}\}\equiv\{\mathcal{F}_{n_{c}}|n_{c}=1,2,...,N-1\}$ of the threshold fidelities can serve as a set of graduations to indicate the degree of network imperfection. That is, if the measured fidelity $F$ is found to be $\mathcal{F}_{n'_{c}+1}<F\leq\mathcal{F}_{n'_{c}}$, then one can infer that there are $n'_{c}$ classical nodes in the created network. Note that the preexisting state model used here is distinct from hidden variable models, such as the Mermin-Peres square, where noncontextual outcomes apply to each of nine observables for the tests of state-independent quantum contextuality in two-qubit systems \cite{Mermin90,Peres90,Mermin93,Cabello08,Amselem09,kirchmair09}. By contrast, our quantum-classical hybrid model for the derivation of $\mathcal{F}_{n_{c}}$ combines both preexisting outcomes from $n_c$ classical nodes and quantum measurements performed in $n_q$ quantum nodes.

Indeed, the collection $\{\mathcal{F}_{n_{c}}\}$ quantitatively describes how the nonclassical correlations among nodes of the graph states vary between the quantum-classical hybrids. If $F>\mathcal{F}_{n_{c}}$ for a created network, then it is impossible to simulate the correlations between nodes using any networks mixed with classical defects of the minimum classical nodes, $n_c$. Such a quantum characteristic can be interpreted as the genuine multi-subsystem EPR steering \cite{SM}, a new type of genuine multipartite EPR steerability \cite{He13} of graph states \cite{Li15}. Notably, the new-found criterion {$F>\mathcal{F}_{n_{c}}$} is stricter than EW ${F}>1/2$ for genuine multipartite entanglement \cite{Horodecki09,Guhne09,McCutcheon16} of networks, in which a network containing classical nodes can mimic the networks with genuine multipartite entanglement to show $1/2<{F}<\mathcal{F}_{1}$. This serious flaw makes EW unreliable in verification of genuine multipartite entanglement for distributed quantum tasks.

\begin{figure*}[t!]
\centering
\includegraphics[width=1.8\columnwidth]{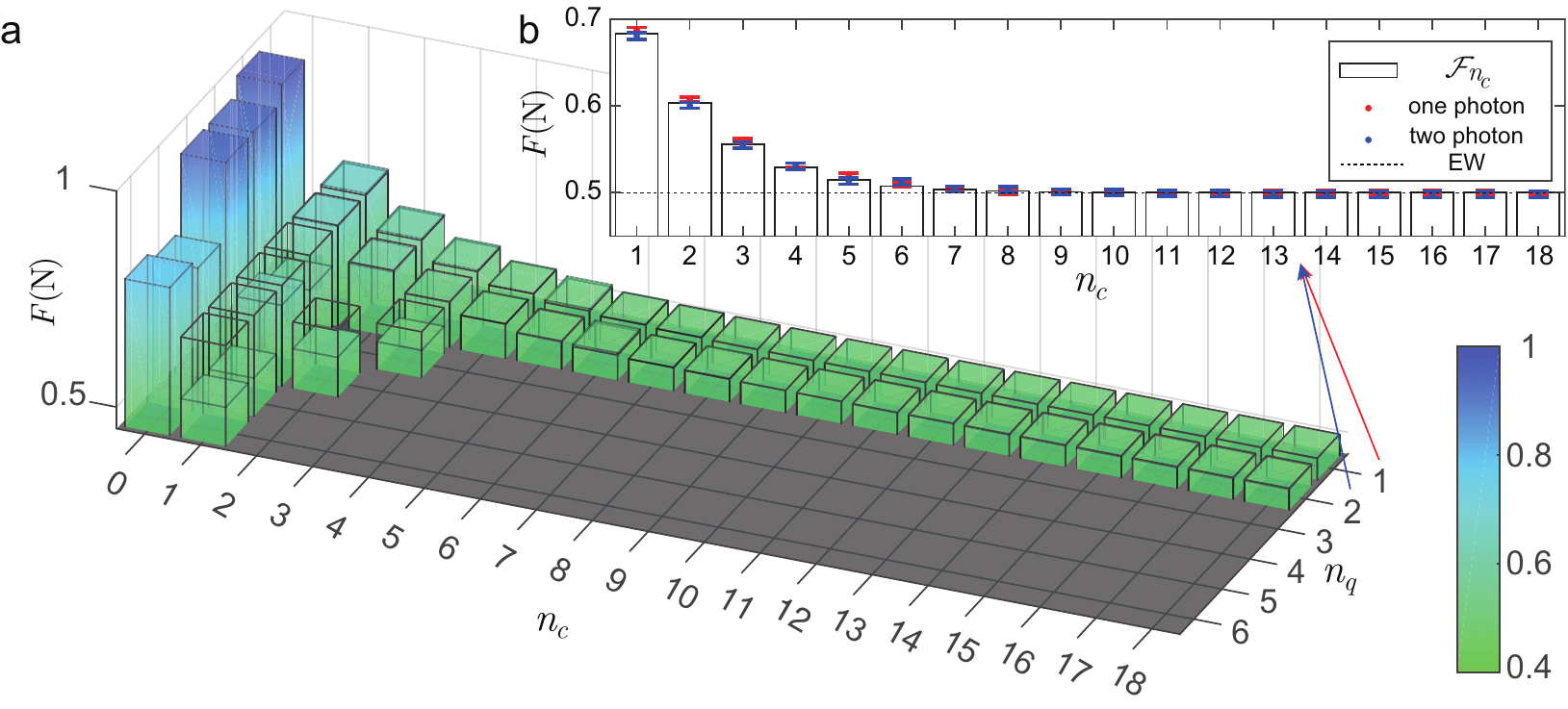}
\caption{Experimental results of network fidelity $F(N)$ in network with variant quantum and classical node number $n_q$ and $n_c$, respectively. (a) The bars edged with black line represent threshold of network fidelity $\mathcal F_{n_c}$. The filled color bars represent the experimental measured network fidelity $F(N)$ in its corresponding network. (b) The results of $F(N)$ in networks with $n_q=1$ and $n_q=2$. The bars represent $\mathcal{F}_{n_c}$, and red (blue) dots represent the measured $F(N)$ in the network with $n_q=1$ ($n_q=2$). \textcolor{blue}{ }}
\label{fig:data2}
\end{figure*}

%\begin{figure}[t!]
%\centering
%\includegraphics[width=1.8\columnwidth]{significanceV2}
%\caption{
%Experimental results of statistical significance $\mathcal S(EW)$ and $\mathcal S(n_c)$ for witnessing entanglement and counting classical nodes, respectively. (a) $\mathcal S(EW)$ and (b) $\mathcal S(n_c)$ in quantum networks with $n_q=1$. (c) $\mathcal S(EW)$ and (d) $\mathcal S(n_c)$ in quantum networks with $n_q=2$. The cyan bars represent positive deviation and magenta bars represent negative deviation. 
%The insets are $\mathcal S(EW)$ and $\mathcal S(n_c)$ from $n_c=8$ to $n_c=18$ in enlarged scale.
%}
%\label{fig:data3}
%\end{figure*}

We experimentally demonstrate our protocol on multipartite graph states in star graph $\ket{G^{star}_N}$, which is equivalent to a Greenberg-Horne-Zeilinger (GHZ) state $\ket{GHZ}_N=(1/\sqrt{2})(\ket{0}^{\otimes N}+\ket{1}^{\otimes N})$ via local operation and classical communication (LOCC). The experimental setup to generate a six-photon GHZ state $\ket{GHZ}_6=(1/\sqrt{2})(\ket{H}^{\otimes 6}+\ket{V}^{\otimes 6})$ with $H$ the horizontal polarization and $V$ the vertical polarization is shown in Fig.~\ref{fig:setup}(b). An experimental state, denoted as $\rho^{GHZ}_6$, is generated by employing the typical spontaneous parametric down-conversion entangled photon source and photonic interferometry technologies (see SM for more details \cite{SM}). The network fidelity $F(6)$ of the generated state $\rho^{GHZ}_6$ is measured by the device shown in Fig.~\ref{fig:setup}(c), which is consisted of a quarter-wave plate (QWP), half-wave plate (HWP), a polarization beam splitter (PBS) and two detectors. By properly choosing the angle of QWP and HWP, the expected value of $I$, $X$, $Y$ and $Z$ can be readout. The experimental results of measured $F(6)$ are shown in Fig.~\ref{fig:data1}(a), from which we calculate that $F(6)=0.792\pm0.006$ [shown with blue bar in Fig.~\ref{fig:data1}(c)]. Then $F(6)$ exceeds the threshold fidelity $\mathcal F_1=0.683$ by more than 18 standard deviation, which indicates there is no classical node in the tested network.   

We then consider the case where $n_c$ classical nodes exist in the $N$-node network, and show that $F(N)$ is bounded by the threshold fidelity $\mathcal F_{n_c}$ even with the optimal ``cheating strategy'' (OCS). The OCS of one untrusted node in bipartite quantum correlation has been well discussed \cite{Bennet12}. We generalize the OCS of $n_c$ untrusted (classical) nodes existing in $N$-node network as: the $n_c$ untrusted nodes first prepare the entangled state $\ket{\xi}_{n_q}$ for $n_q$ trusted (quantum) nodes based on their knowledge of the $N$-node network, where $n_q=N-n_c$. Then, according to the measurement setting for the network fidelity function [Eq.~(\ref{fg})], the $n_c$ untrusted nodes broadcast their results $\in\{+1, -1\}$ to achieve the maximal $F(N)$ \cite{SM}. The experimental results of $F(6)$ under the OCS are shown in Fig.~\ref{fig:data1}(b), from which we calculate $F(6)=0.538\pm0.007$ [shown with red bar in Fig.~\ref{fig:data1}(c)]. As shown in Fig.~\ref{fig:data2}(a), we can see that $F(6)$ either with $n_c=0$ or $n_c=1$ exceeds the EW threshold \cite{McCutcheon16}, which is a strong evidence that EW is no longer reliable in quantum network identification. However, with our criteria, the measured $F(6)$ does not exceed the threshold fidelity $\mathcal F_1$, which indicates there are classical nodes in the measured network. One may notice that, in the case of $n_c=1$ under the OCS, $F(6)=0.538\pm0.007$ is lower than $\mathcal F_3$ but higher than $\mathcal F_4$, where the fidelity threshold overcounts the number of classical nodes in the created network. This is caused by the imperfections in the state preparation, where $F(6)=0.792\pm0.006$ in the case of $n_c=0$, and such imperfections prevent us from achieving the optimal fidelity $\mathcal{F}_{1}\simeq0.6830$ with the OCS. These imperfections can be evaluated by our fidelity criteria in terms of the number of classical nodes. For $\mathcal{F}_{4}<F(6)=0.538\pm0.007<\mathcal{F}_{3}$, the quantity of classical defects in the created network effectively equals to three classical nodes with the optimal mimicry.

We also experimentally prepare various $N$-node quantum networks with $n_q$ up to 6 and $n_c$ up to 18, where $N=n_q+n_c$ \cite{SM}. For each network, $n_c$ nodes employ OCS to achieve maximal network fidelity $F(N)$. The measured $F(N)$ are shown in Fig.~\ref{fig:data2}(a). It is clear that $F(N)$ with $n_c$ classical nodes are bounded by the threshold fidelity $\mathcal F_{n_c}$. One may notice that $F(N)$ decreases much faster than $\mathcal F_{n_c}$ as $n_q$ is increased. This is mainly caused by the imperfections in the state preparation, in which more imperfections are introduced when coherently manipulating more photons. When $n_q\geq3$, $F(N)$ decreases below EW threshold (0.5) quickly as $n_c$ is increased ($n_c\geq4$). We investigate $F(N)$ of networks with $n_q=1$ and  $n_q=2$ for large $n_c$ as the one-photon and two-photon states are prepared with high fidelities. The results of $F(N)$ for $n_q=1$ and  $n_q=2$ are particularly shown in Fig.~\ref{fig:data2}(b), from which we can see that $F(N)$ fits $\mathcal F_{n_c}$ very well. We analyze the standard deviation $\mathcal E$ of $F(N)$ in verifying entanglement and evaluating $n_c$. We observe $\mathcal E>3$ when $n_c\leq 6$ in the created network with $n_q=1$ and $n_q=2$, which reflects a high confidence level of our criteria \cite{SM}.
 
Our results, to the best of our knowledge, demonstrate the first method capable of counting the number of classical nodes in quantum networks. Moreover, the proposed method reveals that the quantum-classical hybrid networks with OCS can surpass the seminal EW threshold of ${F}>1/2$, which causes serious flaws in using the verification of EW in quantum networks \cite{McCutcheon16}. Our proof-of-principle photonic networking experiments, with $n_q$ up to 6 and $n_c$ up to 18, validated the proposed threshold network fidelities $\mathcal{F}_{n_{c}}$, and showed the failure of using EW for genuine multipartite entanglement verification. Our results therefore not only open a new way to characterize classical defects in quantum networks \cite{Kimble08,Wehner18,Ritter12,Northup14,Monroe14,Reiserer15,Sipahigil16,Kalb17,Pirker17,Humphreys18,Chou18,Jing19,Yu20} for a wide range of distributed quantum tasks \cite{Hillery99,Chen05,Markham08,Bell14,Lu16,Huang19,Komar14,Proctor18,Raussendorf01,Walther05,Broadbent09,Barz12,Chen07,Lo14,Epping17}, but also provide novel insights in multipartite nonclassical correlations in graph states \cite{Monz11,Aguilar14,Aolita15}. We expect that our formalism could be extended to the other types of quantum states, such as $W$ states, for characterization of multipartite entanglement with further studies \cite{Horodecki09,Guhne09,Walter09}.

\bibliography{CCQN_revised}

\clearpage

\section*{Supplementary Information}

\section*{Graph states}

Suppose that each quantum node is a quantum two-dimensional system (qubit). An edge, say $(i,j)\in E$, corresponds to a two-qubit conditional transformation among the two qubits (vertices) $i$ and $j$ by $U_{(i,j)}\!\!=\!\!\sum_{v_{i}=0}^{1}\left|v_{i}\right\rangle\!\left\langle v_{i}\right|\otimes (Z_{j})^{v_{i}}$, where $\{\left|v_{i}\right\rangle\}$ is an orthonormal basis of the $i$th qubit and $Z_{j}=\sum_{v_{j}=0}^{1}(-1)^{v_{j}}\left|v_{j}\right\rangle\!\left\langle v_{j}\right|$. The state vector of a graph state is determined by the target graph $G(V,E)$, where $|V|=N$ indicates the total node number and $E$ tells us how the nodes are connected together to show the network topology; that is, 
\begin{equation}
\left|G\right\rangle=\prod_{(i,j)\in E}U_{(i,j)}\left|f_{0}\right\rangle,\label{G}
\end{equation}
where $\left|f_{0}\right\rangle=[(\left|0\right\rangle+\left|1\right\rangle)/\sqrt{2}]^{\otimes N}$ is the initial state of the network.

\section*{Fidelity function $F(N)$}
For a $N$-node network in arbitrary graph state $\left|G\right\rangle$ as shown in the main text, the value of the fidelity function $F(N)$ under given experimental results shows the fidelity of the created network and the target graph state. The way the graph state $\left|G\right\rangle\!\!\left\langle G\right|$ is decomposed decides the construction of the fidelity function, which can be understood by the following explicit general decomposition for arbitrary graph states \cite{Nielsen10,James01}
\begin{equation}
\left|G\right\rangle\!\!\left\langle G\right|=\sum_{\vec{m}}h_{\vec{m}}\bigotimes_{k=1}^{N}\hat{R}_{m_{k}},\label{qst}
\end{equation}
where $\hat{R}_{m_{k}}$ represents the $m_{k}$th observable of the $k$th qubit. For instance, suppose that the created network is in a $N$-qubit state described by a density operator $\rho_{\text{expt}}$, the network fidelity function of $\rho_{\text{expt}}$ and $\left|G\right\rangle\!\!\left\langle G\right|$ satisfies the relation
\begin{eqnarray}
F(N)&=&\text{tr}(\rho_{\text{expt}}\left|G\right\rangle\!\!\left\langle G\right|)\nonumber\\
&=&\sum_{\vec{m}}h_{\vec{m}}\left\langle R_{m_{1}}...R_{m_{N}}\right\rangle,\label{Function}
\end{eqnarray}
where $\left\langle R_{m_{1}}...R_{m_{N}}\right\rangle=\text{tr}(\rho_{\text{expt}}\bigotimes_{k=1}^{N}\hat{R}_{m_{k}})$.

In this work, we assume that the measurements on each qubit are performed with the observables in the Pauli matrices, $\{\hat{R}_{m_{k}}|m_{k}=0,1,2,3\}$, where $\hat{R}_{0}=I$, $\hat{R}_{1}=X$, $\hat{R}_{2}=Y$, and $\hat{R}_{3}=Z$. The spectral decomposition of the Pauli matrices: $\hat{R}_{0}=\sum_{v_{m_{k}}=\pm 1}\left|v_{m_{k}}\right\rangle_{m_{k}m_{k}}\!\!\left\langle v_{m_{k}}\right|$, and $\hat{R}_{m_{k}}=\sum_{v_{m_{k}}=\pm 1}v_{m_{k}}\left|v_{m_{k}}\right\rangle_{m_{k}m_{k}}\!\!\left\langle v_{m_{k}}\right|$ for $m_{k}=1,2,3$, reminds us the relation between the measurement outcomes $R_{0}=\text{tr}(\left|v_{m_{k}}\right\rangle_{m_{k}m_{k}}\!\!\left\langle v_{m_{k}}\right|\hat{R}_{0})=1$ and $R_{m_{k}}=\text{tr}(\left|v_{m_{k}}\right\rangle_{m_{k}m_{k}}\!\!\left\langle v_{m_{k}}\right|\hat{R}_{m_{k}})=v_{m_{k}}$ and the state of the measured qubit $\left|v_{m_{k}}\right\rangle_{m_{k}}$. Note that there are the following relationships between the different eigenstates of $\hat{R}_{m_{k}}$: $\left|\pm1\right\rangle_{1}=(\left|0\right\rangle\pm\left|1\right\rangle)/\sqrt{2}$ and $\left|\pm1\right\rangle_{2}=(\left|0\right\rangle\pm i\left|1\right\rangle)/\sqrt{2}$, where $\left|0\right\rangle\equiv\left|1\right\rangle_{3}$ and $\left|1\right\rangle\equiv\left|-1\right\rangle_{3}$ are also used in the definition of graph state~(\ref{G}). Therefore the states $\left|G\right\rangle\!\!\left\langle G\right|$ can be specified by the decomposition (\ref{qst}) using the orthonormal set of matrices, $\{\bigotimes_{k=1}^{N}\hat{R}_{m_{k}}/\sqrt{2}\}$, from which the fidelity function $\mathcal{F}$ is then constructed. In this case the constituent $2^{N}$ matrices with $h_{\vec{m}}=2^{-N}$ consist of the stabilizer of the graph state \cite{Hein04}. In addition to Pauli matrices used here, the fidelity function can also be constructed in the same manner when the observables for state decomposition are not orthonormal.\\

\section*{Classical nodes}
Suppose a given $N$-node network with the desired target graph state decays into a hybrid of quantum and classical nodes. As shown in the main text, the index set of the network nodes, $V$, can then be divided into the quantum-node subset, $V_{Q}$, and the classical-node subset, $V_{c}$, accordingly. We assume that there are $n_{c}$ classical nodes in the hybrid network, i.e., $|V_{c}|=n_{c}$, and $|V_{c}|+|V_{Q}|=N$. Classical nodes possess physical properties that exist independent of observation \cite{Guhne05}. The state of each classical node can be specified by a preexisting and fixed set of measurement outcomes \cite{Li15}
\begin{equation}
\textbf{v}_{k}\equiv\{R_{m_{k}}=v_{m_{k}}|m_{k}=1,2,3\},\label{cvk}
\end{equation}
where $k\in V_{c}$. For $v_{m_{k}}=\pm1$, we have $8$ possible such sets denoted by $\textbf{v}_{k,\eta}=\{v_{1},v_{2},v_{3}\}$, where
\begin{eqnarray}
&&\textbf{v}_{k,1}=\{+1,+1,+1\}, \textbf{v}_{k,2}=\{+1,+1,-1\}, \nonumber \\
&&\textbf{v}_{k,3}=\{+1,-1,+1\}, \textbf{v}_{k,4}=\{+1,-1,-1\}, \nonumber \\
&&\textbf{v}_{k,5}=\{-1,+1,+1\}, \textbf{v}_{k,6}=\{-1,+1,-1\}, \nonumber \\
&&\textbf{v}_{k,7}=\{-1,-1,+1\}, \textbf{v}_{k,8}=\{-1,-1,-1\}.\nonumber 
\end{eqnarray}
Incoherent manipulations of qubits cause the qubits to decay such that the qubit states can eventually be described by the classical realistic theory, such as measurements on qubits, and qubit storage or transmission under the action of a noisy channel.

For example, suppose that the observable $Z$ is chosen for a measurement on the superposition state $\ket{1}_{1}=(\ket{1}_{3}+\ket{-1}_{3})/\sqrt{2}$. After measurement, the qubit becomes a definite state in one of the two post-measurement states $\ket{1}_{3}$ and $\ket{-1}_{3}$, independent of the observation with respect to the measurement of the observable $Z$. That is, in terms of the classical realistic theory, if the measurement outcome is $v_{3}=1$ ($v_{3}=-1$), the probability of constantly observing $v_{3}=1$ ($v_{3}=-1$) under the measurement of $Z$ is $p(v_{3}=1)=\sum_{\eta=1,3,5,7}p(\textbf{v}_{k,\eta})=1$ ($p(v_{3}=-1)=\sum_{\eta=2,4,6,8}p(\textbf{v}_{k,\eta})=1$), where $p(\textbf{v}_{k,\eta})$ is the probability of being in the preexisting state $\textbf{v}_{k,\eta}$.

Moreover, after many rounds of the same measurements on the same initial superposition state, and taking all post-measurment states into account, the average post-measurement state, $\rho_{\text{avg}}=(\ket{1}_{33}\!\bra{1}+\ket{-1}_{33}\!\bra{-1})/2$, can be described in the same manner. One can think of $\rho_{\text{avg}}$ as one composed of preexisting states $\textbf{v}_{k,\eta}$ with a probability distribution $p(\textbf{v}_{k,\eta})$ such that $p(v_{3}=1)=\sum_{\eta=1,3,5,7}p(\textbf{v}_{k,\eta})=1/2$ and $p(v_{3}=-1)=\sum_{\eta=2,4,6,8}p(\textbf{v}_{k,\eta})=1/2$.

From the viewpoint of physically motivated operational meaning, once we have the preexisting recipe of $\textbf{v}_{k,\eta}$ with the probability distribution $p(\textbf{v}_{k,\eta})$, the post-measurement state $\rho_{\text{avg}}$ can be prepared accordingly by incoherently mixing the states $\ket{1}_{3}$ and $\ket{-1}_{3}$. It is clear that $\textbf{v}_{k,\eta}$ exist independent of observation; after all, the prescribed information about $\textbf{v}_{k,\eta}$ has been shown in the preexisting recipe.

The descriptions of the classical nodes illustrated above can be directly applied to the cases where qubit storage or transmission undergoes noisy channels, such as the final state $\ket{v_{3}}_{3}$ which results from the initial state $\ket{1}_{1}$ through an amplitude damping channel, and the complete mixed state $\rho_{\text{avg}}$ derived from $\ket{1}_{1}$ via a full decoherence process.

\section*{Set of graduations $\{\mathcal{F}_{n_{c}}\}$}
To examine the minimum deviation of a $N$-node network with $n_{c}$ classical nodes from the target graph state, we evaluate the maximum fidelity by performing the following task
\begin{equation}
\mathcal{F}_{n_{c}}=\max_{V_{c},\{R_{m_{k}}\}}F,\label{Fnc}
\end{equation}
where the maximization is over all vertex sets $V_{c}$ with $|V_{c}|=n_{c}$ and all outcomes from measurements on the classical and quantum nodes, $\{R_{m_{k}}\}$. Next, we illustrate the step-by-step derivation for Eq.~(3-6) in the main text. The fidelity (\ref{Fnc}) is determined in three steps:

First, as shown in Eq.~(3) in the main text, before the state decay, when explicitly considering Eq.~(\ref{Fnc}), the state vector of the target graph state can always be represented in the Schmidt form of rank $r$ \cite{Li10}
\begin{equation}
\left|G\right\rangle=\frac{1}{\sqrt{r}}\sum_{v=0}^{r-1}\left|v\right\rangle_{sQ}\left|v\right\rangle_{sc},\label{SD}
\end{equation}
for $r\geq2$, where $\{\left|v\right\rangle_{sQ}\}$ and $\{\left|v\right\rangle_{sc}\}$ are the Schmidt bases for the nodes in the vertex sets $V_{Q}$ and $V_{c}$, respectively. This representation shows us the following form under the state decomposition for the nodes in $V_{c}$
\begin{equation}
\left|G\right\rangle\!\!\left\langle G\right|=\frac{1}{r}\sum_{v,v',\vec{m}_{c}}h_{\vec{m}_{c}}^{vv'}\left|v\right\rangle_{\!sQsQ}\!\!\left\langle v'\right|\bigotimes_{k\in V_{c}}\hat{R}_{m_{k}}, \label{StDecom}
\end{equation}
where $\left|v\right\rangle_{\!scsc}\!\!\left\langle v'\right|=\sum_{\vec{m}_{c}}h_{\vec{m}_{c}}^{vv'}\bigotimes_{k\in V_{c}}\hat{R}_{m_{k}}$, $\vec{m}_{c}\equiv \{m_{k}|k\in V_{c}\}$, and $h_{\vec{m}_{c}}^{vv'}$ denote the decomposition coefficients. Alternatively, it can be represented in the matrix form in the basis, $\{\left|v\right\rangle_{sQ}\}$. Let us take $r=2$ for example, we have 
\begin{equation}
\left|G\right\rangle\!\!\left\langle G\right|=\left(\begin{array}{cc}\hat{f}_{00} & \hat{f}_{01} \\ \hat{f}_{10} & \hat{f}_{11}\end{array}\right), \label{StDecom2}
\end{equation}
where $\hat{f}_{vv'}=1/2\sum_{\vec{m}_{c}}h_{\vec{m}_{c}}^{vv'}\bigotimes_{k\in V_{c}}\hat{R}_{m_{k}}$.

Second, we use the classical realistic theory to give a complete description of the total state of the $n_{c}$ classical nodes in terms of the preexisting outcome sets: $\{\textbf{v}_{k}|k\in V_{c}\}$. The network fidelity function (Function) can be rephrased as the following explicit form
\begin{equation}
F=\frac{1}{r}\sum_{v,v',\vec{m}_{c}}h_{\vec{m}_{c}}^{vv'}\left\langle\left|v\right\rangle_{\!sQsQ}\!\!\left\langle v'\right|\right\rangle\left\langle\prod_{k\in V_{c}}R_{m_{k}}\right\rangle.\label{F}
\end{equation}
Note that $R_{m_{k}}$ are the preexisting outcomes given in $\textbf{v}_{k}$ (\ref{cvk}) where $\textbf{v}_{k}\in\{\textbf{v}_{k,\eta}|\eta=1,2,...,8\}$.

Finally, through the expression (\ref{F}) for the fidelity function, the maximization task (\ref{Fnc}) becomes
\begin{equation}
\mathcal{F}_{n_{c}}=\max_{\{\textbf{v}_{k}|k\in V_{c}\}}E[\left(\begin{array}{cc} f_{00} & f_{01} \\ f_{10} & f_{11}\end{array}\right)],\label{Fnc2}
\end{equation}
where $f_{vv'}=1/2\sum_{\vec{m}_{c}}h_{\vec{m}_{c}}^{vv'}\left\langle\prod_{k\in V_{c}}R_{m_{k}}\right\rangle$ and $E[\cdot]$ denotes the largest eigenvalue of the matrix represented in the orthonormal basis $\{\left|0\right\rangle_{sQ},\left|1\right\rangle_{sQ}\}$. Note that, given a set of $f_{vv'}$, the eigenvector of $E[\cdot]$ correspond to the state of quantum nodes. Here $r=2$ is shown to be necessary for the maximum of $F$. One always can find at least one bipartite splitting of the network nodes in the target graph to have such Schmidt rank of state decomposition under the condition $|V_{c}|=n_{c}$. When the fidelity functions are specified in the orthonormal sets of Pauli matrices and the classical nodes are described under the assumption of realism \cite{Guhne05} given in (\ref{cvk}), we obtain 
\begin{eqnarray}
&&f_{00}=\frac{1}{2}, f_{01}=\frac{1}{2^{n_{c}+1}}(1+i)^{n_{c}},\nonumber\\
&&f_{10}=\frac{1}{2^{n_{c}+1}}(1-i)^{n_{c}}, f_{11}=0. \label{fmn}
\end{eqnarray}
Thus we arrive the result of Eq.~(6) in the main text by calculating the maximum eigenvalue of the matrix with the above matrix elements. From the viewpoint of the operational meaning, as the quantum nodes are prepared in the eigenstate of such maximum eigenvalue and the classical nodes are in a specific state $\{\textbf{v}_{k}|k\in V_{c}\}$, the resulting hybrid of quantum and classical nodes can show the best fidelity, $\mathcal{F}_{n_{c}}$.

It is worth noting that the choice of $\{\textbf{v}_{k}|k\in V_{c}\}$ for the maximum eigenvalue $\mathcal{F}_{n_{c}}$ is not unique. Then there exist alternative to the matrix elements $f_{vv'}$ (\ref{fmn}) for matrices with the same maximum eigenvalue, $\mathcal{F}_{n_{c}}$, but different eigenvectors. This also means that there are more than one quantum-classical hybrids that can be used to achieve the best fidelity, $\mathcal{F}_{n_{c}}$.

Next, we give a concrete example of three-qubit star-graph (or chain-graph) state to illustrate the three steps in our method for determining $\mathcal{F}_{n_{c}}$ (\ref{Fnc}). According to (\ref{G}) and (\ref{SD}) a three-qubit star-graph state can be expressed as
\begin{equation}
\left|G_{3}^{star}\right\rangle=\frac{1}{\sqrt{2}}(\left|0\right\rangle_{sQ}\left|0\right\rangle_{sc}+\left|1\right\rangle_{sQ}\left|1\right\rangle_{sc}). \nonumber
\end{equation}
As will be shown bellow, the Schmidt basis $\{\ket{0}_{sc},\ket{1}_{sc}\}$ corresponds to either one ($n_{c}=1$) or two ($n_{c}=2$) nodes in $V_{c}$.

In the case of $n_c=1$, we assume that the 3rd qubit is the node belonging to $V_{c}$ and connected with the other two qubits in $V_{Q}$. Then we have 
\begin{equation}
\{\left|0\right\rangle_{sQ}=\left|1\right\rangle_{1}\left|1\right\rangle_{1}, \left|1\right\rangle_{sQ}=\left|-1\right\rangle_{1}\left|-1\right\rangle_{1}\},\nonumber
\end{equation}
and
\begin{equation}
\{\left|0\right\rangle_{sc}=\left|1\right\rangle_{3}, \left|1\right\rangle_{sc}=\left|-1\right\rangle_{3}\}.\nonumber
\end{equation}
With the following explicit decomposition for the nodes in $V_{c}$:
\begin{eqnarray}
&&\left|0\right\rangle_{\!scsc}\!\!\left\langle 0\right|= \frac{1}{2}(\hat{R}_{0}+\hat{R}_{3}), \left|0\right\rangle_{\!scsc}\!\!\left\langle 1\right|= \frac{1}{2}(\hat{R}_{1}+i\hat{R}_{2}),\nonumber\\ 
&&\left|1\right\rangle_{\!scsc}\!\!\left\langle 0\right|= \frac{1}{2}(\hat{R}_{1}-i\hat{R}_{2}), \left|1\right\rangle_{\!scsc}\!\!\left\langle 1\right|= \frac{1}{2}(\hat{R}_{0}-\hat{R}_{3}),\nonumber
\end{eqnarray}
we obtain Eq.~(\ref{StDecom}). As the 3rd qubit decays to a classical node which can be specified by a preexisting state $\textbf{v}_{3}$, any observable in the classical node has a predetermined value that exists independent of observation according to (\ref{cvk}), and from which we arrive at Eq.~(\ref{F}). When setting $\textbf{v}_{3}$ as $\textbf{v}_{3,1}$, we obtain the results of Eq.~(\ref{fmn}) with
\begin{equation}
f_{00}=\frac{1}{2}, f_{01}=\frac{1}{4}(1+i), f_{10}=\frac{1}{4}(1-i), f_{11}=0,\nonumber
\end{equation}
and the minimum deviation of a network mixed with a single classical node:
\begin{equation}
\mathcal{F}_{1}\simeq0.6830,\nonumber
\end{equation}
as shown in Eq.~(2) for $n_{c}=1$ in the main text. %red

Similarly, in the case of $n_c=2$, where the 3rd qubit is assumed to be in $V_{Q}$ and connected with the other two qubits in $V_{c}$, we have
\begin{equation}
\{\left|0\right\rangle_{sQ}=\left|1\right\rangle_{3}, \left|1\right\rangle_{sQ}=\left|-1\right\rangle_{3}\},\nonumber
\end{equation}
and
\begin{equation}
\{\left|0\right\rangle_{sc}=\left|1\right\rangle_{1}\left|1\right\rangle_{1}, \left|1\right\rangle_{sc}=\left|-1\right\rangle_{1}\left|-1\right\rangle_{1}\}.\nonumber
\end{equation}
To derive Eq.~(\ref{StDecom}), the following state decomposition is used
\begin{eqnarray}
&&\left|0\right\rangle_{\!scsc}\!\!\left\langle 0\right|= \frac{1}{4}(\hat{R}_{0}\hat{R}_{0}+\hat{R}_{0}\hat{R}_{1}+\hat{R}_{1}\hat{R}_{0}+\hat{R}_{1}\hat{R}_{1}),\nonumber\\
&&\left|0\right\rangle_{\!scsc}\!\!\left\langle 1\right|= \frac{1}{4}(-\hat{R}_{2}\hat{R}_{2}-i\hat{R}_{2}\hat{R}_{3}-i\hat{R}_{3}\hat{R}_{2}+\hat{R}_{3}\hat{R}_{3}),\nonumber\\
&&\left|1\right\rangle_{\!scsc}\!\!\left\langle 0\right|= \frac{1}{4}(-\hat{R}_{2}\hat{R}_{2}+i\hat{R}_{2}\hat{R}_{3}+i\hat{R}_{3}\hat{R}_{2}+\hat{R}_{3}\hat{R}_{3}),\nonumber\\
&&\left|1\right\rangle_{\!scsc}\!\!\left\langle 1\right|= \frac{1}{4}(\hat{R}_{0}\hat{R}_{0}-\hat{R}_{0}\hat{R}_{1}-\hat{R}_{1}\hat{R}_{0}+\hat{R}_{1}\hat{R}_{1}).\nonumber
\end{eqnarray}
When the 1st and the 2nd qubits decay into two classical nodes which can be specified by the preexisting states of $\textbf{v}_{1}$ and $\textbf{v}_{2}$, respectively, the fidelity function (\ref{Function}) becomes Eq.~(\ref{F}), and any observable in the two classical nodes has a predetermined value that exists independent of observation according to (\ref{cvk}). As $\textbf{v}_{1}=\textbf{v}_{1,3}$ and $\textbf{v}_{2}=\textbf{v}_{2,3}$, we obtain the elements of the matrix represented in the orthonormal basis [see Eqs.~(\ref{Fnc2}) and (\ref{fmn})]:
\begin{equation}
f_{00}=\frac{1}{2}, f_{01}=\frac{i}{4}, f_{10}=-\frac{i}{4}, f_{11}=0, \nonumber
\end{equation}
and from which we get the the maximum fidelity for a network consisting of one quantum node and two classical nodes:
\begin{equation}
\mathcal{F}_{2}\simeq0.6036.\nonumber
\end{equation}
This result consistent with Eq.~(2) for $n_{c}=2$ shown in the main text.

\begin{figure*}[t!]
\centering
\includegraphics[width=2\columnwidth]{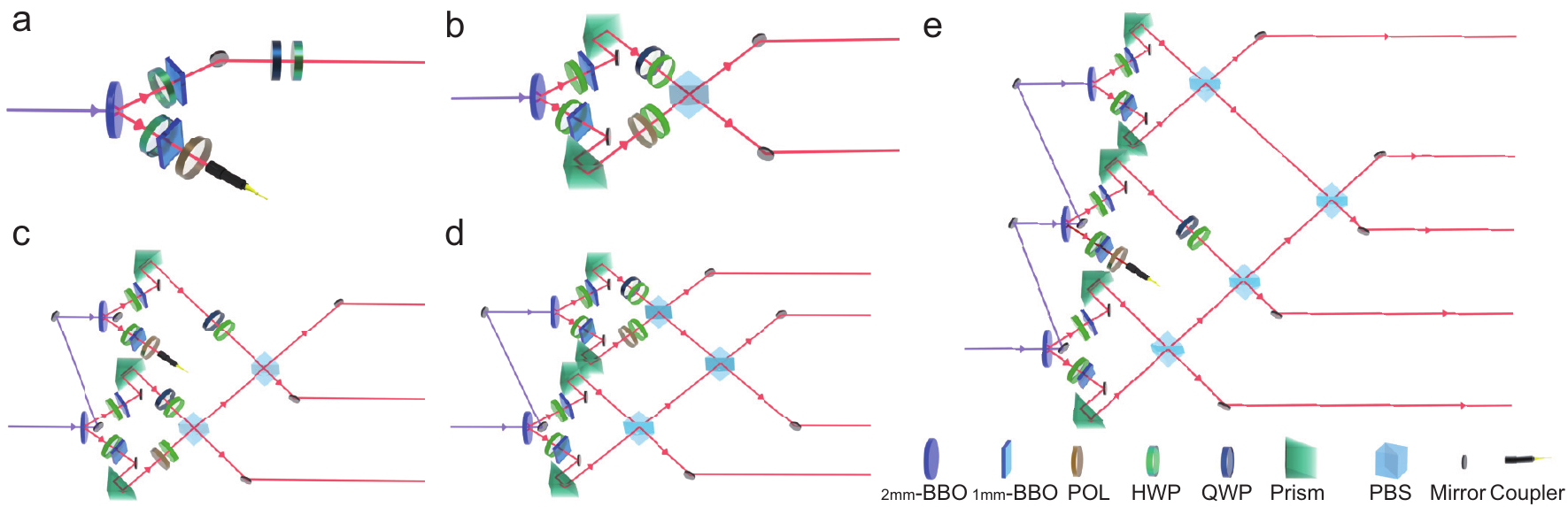}
\caption{Experimental setups for realizing quantum networks with $n_q+n_c$ nodes. (a-e) setup to implement networks of $n_q=1$, $n_q=2$, $n_q=3$, $n_q=4$ and $n_q=5$, respectively.}
\label{fig:setup2}
\end{figure*}

\section*{Genuine multi-subsystem EPR steering}
Satisfying the criterion
\begin{equation}
F>\mathcal{F}_{n_{c}},\label{FSteering}
\end{equation}
confirms that, the correlation between the network nodes of a created state is stronger than all the correlations that can be created by the quantum-node subsets, $V_{Q}$, and the classical-node subsets, $V_{c}$ with $|V_{c}|=n_{c}$, for all possible bipartitions of the hybrid of quantum and classical nodes. This concretely describes the steering effects \cite{Wiseman07,Jones07} between two subsystems with $n_{c}$ and $n_{q}$ nodes, respectively. Moreover, since all possible configurations of splitting $N$ nodes into two subsystems are considered in the criterion, we call such steerability the genuine multi-subsystem EPR steering. This description generalizes the concept of genuine multipartite EPR steering \cite{He13,Li15}, where only the extreme value of $\mathcal{F}_{n_{c}}$ is involved. Such steerability is shown if the created network with a fidelity that goes beyond the threshold
\begin{equation}
\max_{n_{c},V_{c},\{R_{m_{k}}\}}F=\max_{n_{c}}\mathcal{F}_{n_{c}}.\label{FSteering2}
\end{equation}
From the result of $\mathcal{F}_{n_{c}}$ , it is clear that $\mathcal{F}>\mathcal{F}_{1}$ is a fidelity criterion for genuine multipartite EPR steering.

\begin{figure*}[t!]
\includegraphics[width=\linewidth]{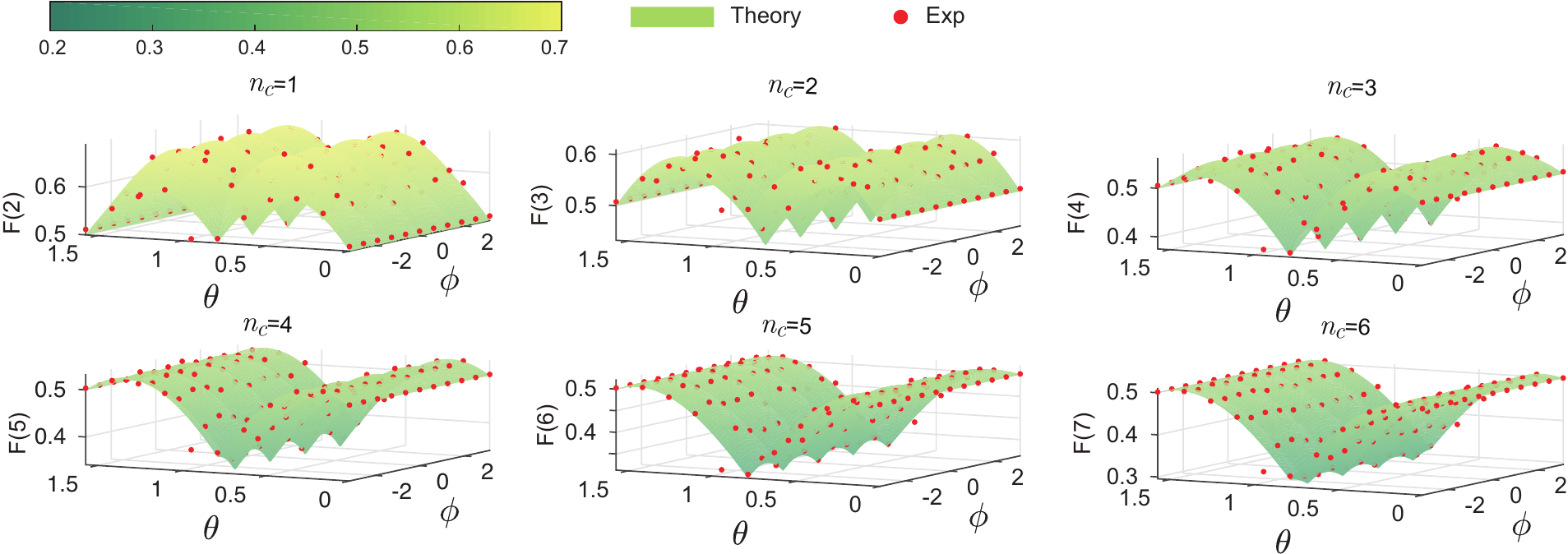}
\caption{Experimental results of network fidelity in the case of $n_q=1$ with different $\theta$ and $\phi$ setting. The landscapes are theoretical results. The experimental results are show with red dots. The errors are shown with red bars, which are too small to recognize in the presented scale. (a)$\sim$(f) show the cases of $n_c=1$ to $n_c=6$, respectively.}\label{fig:surf_Q1Cn}
\end{figure*}

\section*{Optimal Cheating Strategy (OCS) and its experimental realization}

In the $N$-node network with $n_c$ untrusted (classical) nodes and $n_q$ trusted (quantum) nodes, the $n_c$ untrusted nodes prepare the entangled state $\ket{\xi}_{n_q}$ based on their knowledge of the network fidelity function $F(N)$. Combining with the measurement results of prepared $\ket{\xi}_{n_q}$, the $n_c$ untrusted nodes broadcast their results $\in\{+1, -1\}$, according to the measurement setting for the network fidelity function. In this section, we detail the experimental state preparation of quantum networks with $N=n_c+n_q$ nodes.

\subsection{$n_c=0$}

As we mentioned in the main text, we test our procedure on multiphoton graph state in star graph, which is equivalent to N-photon Greenberger-Horne-Zeilinger (GHZ) state via local operation and classical communication (LOCC). The building block to generate a N-photon GHZ state is entangled photon pair from spontaneous parametric down conversion (SPDC). Experimentally, we use an ultraviolet pulse (with central wavelength of 390nm, pulse duration of 140fs) to pump a 2mm Beta Barium Borate (BBO) crystal (as shown in Fig.~\ref{fig:setup2}a). The generated twin photons are entangled both in polarization degree of freedom (DOF) and frequency DOF, and separated with opening angle of 6$^{\circ}$. To compensate the time and spatial walk-off, an half-wave plate set at 45$^{\circ}$ and a 1mm BBO crystal are inserted in each path. Using narrow-band filters with full-width at half maximum ($\lambda_{\text{FWHM}}$) of 3~nm, the maximally entangled photon pair can be obtained $\ket{GHZ}_2=\frac{1}{\sqrt{2}}(\ket{HH}+\ket{VV})$. The generated photons are highly frequency-correlated, which is not suitable for multiphoton experiments. To enhance the multiphoton coincidence count rate, we recombine the generated photons on a polarization beam splitter (PBS) to eliminate frequency correlation\cite{Kim03,Yao12} (similar to the setup shown in Fig.~\ref{fig:setup2}b, but remove polarizer (POL), quarter-wave plate (QWP) and one HWP). By properly choosing narrow bandpass filters (full-width half-maximum (FWHM) of 3nm and 8nm in our experiment), we can obtain $\ket{GHZ}_2$ with higher counter rate. 

To generate $\ket{GHZ}_3$, the ultraviolet pulse goes to shine another 2mm BBO crystal. A heralded single-photon source is obtained by triggering one photon of the second entangled photon pair. The heralded single-photon is rotated to $\ket{+}=\frac{1}{\sqrt{2}}(\ket{H}+\ket{V})$ and then superposed with one photon from the first $\ket{GHZ}_2$ on a PBS. When two photons arrive at PBS simultaneously and come out from two output ports, $\ket{GHZ}_3$ can be obtained. The setup is shown in Fig.~\ref{fig:setup2}c.

$\ket{GHZ}_4$ is generated without triggering one photon from the second SPDC (as shown in Fig.~\ref{fig:setup2}d). Similarly, by successively shinning a third BBO crystal and proper operations, $\ket{GHZ}_5$ and $\ket{GHZ}_6$ can be obtained (as shown in Fig.~\ref{fig:setup2}e and Fig.~1b in the main text).

\subsection{$n_c=1$}

To show the optimal ``cheating strategy" (OCS) for quantum network $\ket{GHZ}_{N}$ with $n_{c}=1$, first, we have the following state decomposition according to Eq. (\ref{StDecom}):
\begin{equation}
\ket{GHZ}_{NN}\!\bra{GHZ}=\frac{1}{2}\sum_{k,l=H,V}\ket{k}^{\otimes N-1\otimes N-1}\!\bra{l}\otimes \ket{k}\!\!\bra{l}\label{GHZN}.
\end{equation}
Alternatively, the above state decomposition can be represented in the matrix form in the basis $\{\ket{H}^{\otimes N-1},\ket{V}^{\otimes N-1}\}$ of the qubits in $V_{Q}$:
\begin{eqnarray}
\ket{GHZ}_{NN}\!\bra{GHZ}&=&\frac{1}{2}\left(\begin{array}{cc}\ket{H}\!\!\bra{H} & \ket{H}\!\!\bra{V} \\\ket{V}\!\!\bra{H} & \ket{V}\!\!\bra{V}\end{array}\right)\nonumber\\
&=&\frac{1}{2}\left(\begin{array}{cc} \frac{1}{2}(I+Z)& \frac{1}{2}(X+iY) \\ \frac{1}{2}(X-iY) & \frac{1}{2}(I-Z)\end{array}\right)\nonumber,
\end{eqnarray}
where $Z\equiv \ket{H}\!\!\bra{H}-\ket{V}\!\!\bra{V}$, $X\equiv \ket{H}\!\!\bra{V}+\ket{V}\!\!\bra{H}$, and $Y\equiv -i\ket{H}\!\!\bra{V}+i\ket{V}\!\!\bra{H}$ for the qubit in $V_{Q}$. Note that the last matrix form corresponds to Eq.~(\ref{StDecom2}).

Secondly, suppose that $\ket{H}\equiv\ket{1}_{3}$ and $\ket{V}\equiv\ket{-1}_{3}$ and apply the assumption of classical state $\textbf{v}_{k,1}=\{+1,+1,+1\}$ to the node in $V_{c}$, the above matrix becomes
\begin{equation}
\left(\begin{array}{cc} \frac{1}{2}& \frac{1}{4}(1+i) \\ \frac{1}{4}(1-i) & 0\end{array}\right),
\end{equation}
where the matrix elements are consistent with the ones shown in Eq. (\ref{fmn}) for $n_{c}=1$. Then, the maximum eigenvalue of this matrix is just the threshold fidelity, $\mathcal{F}_{1}\simeq0.6830$; the corresponding eigenvector is $\ket{\xi}_{N-1}=\cos\theta\ket{H}^{\otimes N-1}+\sin\theta e^{i\phi}\ket{V}^{\otimes N-1}$ of the quantum nodes, where $\sin\theta=\frac{1}{\sqrt{3+\sqrt{3}}}$ and $\phi=-\frac{\pi}{4}$. That is, as the quantum nodes are prepared in $\ket{\xi}_{N-1}$ and the classical node is in the state $\textbf{v}_{k,1}$, the OCS can be achieved to show the best fidelity of $\ket{GHZ}_{N}$ and the hybrid of quantum and classical nodes, $\mathcal{F}_{1}$.

\begin{figure*}[t!]
\centering
\includegraphics[width=1.8\columnwidth]{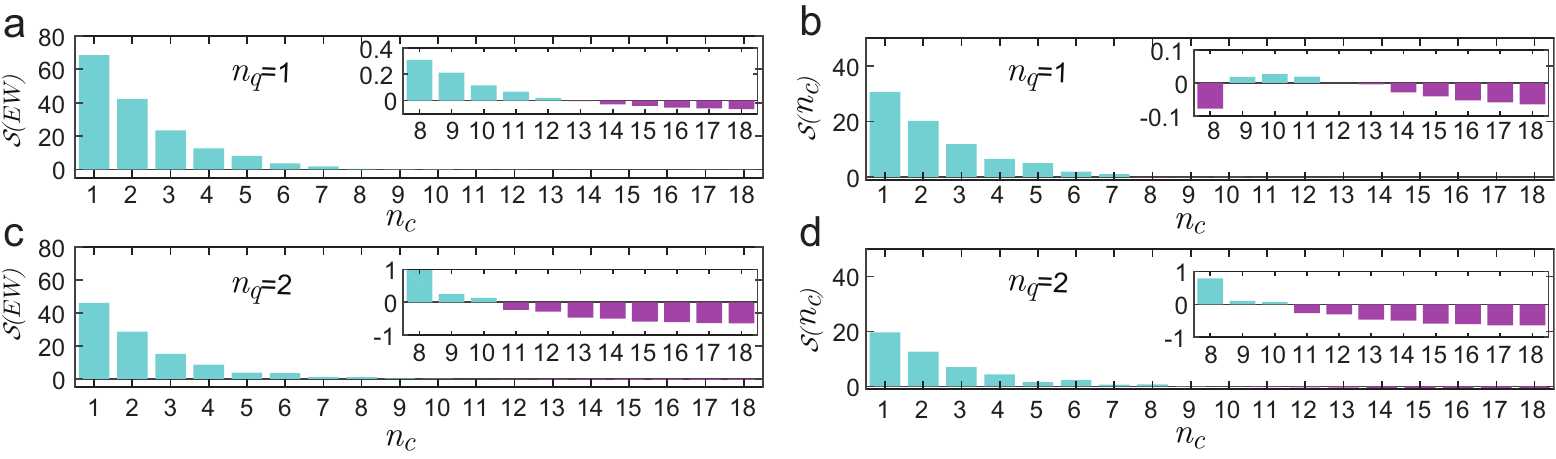}
\caption{
Experimental results of statistical significance $\mathcal S(EW)$ and $\mathcal S(n_c)$ for witnessing entanglement and counting classical nodes, respectively. (a) $\mathcal S(EW)$ and (b) $\mathcal S(n_c)$ in quantum networks with $n_q=1$. (c) $\mathcal S(EW)$ and (d) $\mathcal S(n_c)$ in quantum networks with $n_q=2$. The cyan bars represent positive deviation and magenta bars represent negative deviation. 
The insets are $\mathcal S(EW)$ and $\mathcal S(n_c)$ from $n_c=8$ to $n_c=18$ in enlarged scale.
}
\label{fig:data3}
\end{figure*}

Experimentally, the preparation of $\ket{\xi}_{N-1}$ can be performed by projecting $\ket{GHZ}_{N}$ onto the state $\ket{\xi'}=\cos\theta\ket{H}+\sin\theta e^{-i\phi}\ket{V}$ with QWP and POL shown in Fig.~1d in the main text. By properly choosing the angles of QWP and POL, projection of arbitrary  $\ket{\xi^{\prime}}$ can be realized. As explained in the section of classical nodes, such a measurement can make a quantum node classical, which results in $n_{c}=1$ required in the present cheating scenario.

In the case of $N=6$ with $n_c=1$ and $n_q=5$ shown in Fig.~1 in the main text, we experimentally realize OCS by projecting one photon from $\ket{GHZ}_6$ on $\ket{\xi^{\prime}}=\cos\theta\ket{H}+\sin\theta e^{-i\phi}\ket{V}$ (see Fig.~1d in the main text), which leaves five trusted nodes sharing a five-photon state $\ket{\xi}_5=\cos\theta\ket{H}^{\otimes 5}+\sin\theta e^{i\phi}\ket{V}^{\otimes 5}$. Since the choice of $\textbf{v}_{k}$ for the maximum eigenvalue $\mathcal{F}_{1}$ is not unique, the choice of $\theta$ and $\phi$ in OCS is not unique as well, so that they constitute an ensemble denoted as $\{\ket{\xi}_5\}$. In our experiment, we select $\sin\theta=\frac{1}{\sqrt{3+\sqrt{3}}}$ and $\phi=-\frac{\pi}{4}$ for $\textbf{v}_{k}=\textbf{v}_{k,1}$ as shown above. One can choose other preexisting states to achieve OCS, such as $\textbf{v}_{k}=\textbf{v}_{k,2}$. For such an alternative, one needs to perform the projection onto the state $\ket{\xi^{\prime}}$ with $\sin\theta=\frac{1}{\sqrt{3-\sqrt{3}}}$ and $\phi=-\frac{\pi}{4}$. We also experimentally reconstruct the landscape of network fidelity in the case of $n_q=1$ with different $\theta$ and $\phi$ setting. The results are shown in Fig.~\ref{fig:surf_Q1Cn}

The trusted nodes use the experimental setup shown in Fig.~1c to measure the expected value of $\hat{R}_{m_k}$, in which the transmitted photons are projected on eigenstate $\ket{+1}_{m_k}$ of $\hat{R}_{m_k}$ with eigenvalue $+1$, while the reflected photons are projected on $\ket{-1}_{m_k}$ of $\hat{R}_{m_k}$ with eigenvalue $-1$. The angle setting of QWP and HWP to perform measurements of $X$, $Y$ and $Z$ are shown in Table.~\ref{tab:angle}. Thus, the expected value of $\langle\hat{R}_{m_k}\rangle$ can be obtained by $\frac{N_{+1}-N_{-1}}{N_{+1}+N_{-1}}$, where $N_{+1(-1)}$ is the recorded counts on detector behind transmitted (reflected) photon. For $\hat{R}_{m_k}=I$, the angle setting of QWP and HWP can be arbitrarily chosen, and $\langle I\rangle=\frac{N_{+1}+N_{-1}}{N_{+1}+N_{-1}}=1$. Then, combining the results from untrusted node $\in\{+1, -1\}$ according to the measurement setting for the network fidelity function (1) in the main text, we obtain the network fidelity $F(6)=0.538\pm0.007$, shown with red bar in Fig.~2c in the main text. We emphasize that for the case of $n_c=1$ and $n_q=6$, we choose the OCS of $n_c\geq2$, which means we experimentally prepare a six-photon state to simulate the seven-node network with one photon projected on $\ket{\xi^{\prime}}$.

\begin{table}[h!]
\centering
\begin{tabular}{|c|c|c|c|}
\hline
 Observable & QWP & HWP & Expected value 
\\ \hline
$X$ & 45$^{\circ}$ & 22.5$^{\circ}$ & \multirow{3}{*}{$\frac{N_{+1}-N_{-1}}{N_{+1}+N_{-1}}$} \\
$Y$ & 0$^{\circ}$ & 22.5$^{\circ}$ & \\
$Z$ & 0$^{\circ}$ & 0$^{\circ}$ &   \\
\hline
\end{tabular}
\caption{Angle setting of QWP and HWP in $X$, $Y$, $Z$ detection. \label{tab:angle}}
\end{table}

\subsubsection{$n_c\geq2$}

For general $\ket{GHZ}_N$ with $2\leq n_{c}< N$, we have the following state decomposition according to Eq.~(\ref{StDecom})
\begin{equation}
\ket{GHZ}_{NN}\!\bra{GHZ}=\frac{1}{2}\left(\!\!\begin{array}{cc} \frac{1}{2^{n_c}}(I+Z)^{\otimes n_{c}}& \frac{1}{2^{n_c}}(X+iY)^{\otimes n_{c}} \\ \frac{1}{2^{n_c}}(X-iY)^{\otimes n_{c}} & \frac{1}{2^{n_c}}(I-Z)^{\otimes n_{c}}\end{array}\!\!\!\!\right)\nonumber\!,
\end{equation}
in the basis $\{\ket{H}^{\otimes n_{q}},\ket{V}^{\otimes n_{q}}\}$ of the qubits in $V_{Q}$. When setting $\textbf{v}_{k,1}=\{+1,+1,+1\}$ for $k\in V_{c}$ under the assumption of realism, we get the matrix elements as described in Eq. (\ref{fmn}), and they constitute a matrix with a maximum eigenvalue, $\mathcal{F}_{n_{c}}$, and a corresponding eigenvector, $\ket{\xi}_{n_{q}}=\cos\theta\ket{H}^{\otimes n_{q}}+\sin\theta e^{i\phi}\ket{V}^{\otimes n_{q}}$. Therefore, to achieve OCS for $n_{c}\geq2$, one needs to prepare the state of quantum nodes in $\ket{\xi}_{n_{q}}$ and all the classical nodes in $\textbf{v}_{k,1}$. This corresponds to the choice of
\begin{equation}
\sin\theta=\frac{1}{\sqrt{2^{n_c-1}+2+2^{\frac{-n_{c}}{2}}\sqrt{4+2^{n_{c}}}}},\nonumber
\end{equation}
and $\phi=-\frac{n_c\pi}{4}$, for achieving $\mathcal F_{n_c}$ in our OCS experiments.

Experimentally, we demonstrate the OCS without manipulating $\ket{GHZ}_N$ when $n_c\geq2$, but simulate the networks with $n_c\geq2$ by directly creating the state $\ket{\xi}_{n_q}$ instead. Recall the general OCS for arbitrary $n_c$: the $n_c$ untrusted nodes prepare the entangled state $\ket{\xi}_{n_q}$ for $n_q$ trusted nodes based on their knowledge of the network fidelity function (1) in the main text. We prepare state $\ket{\xi}_{n_q}$ with chosen $\theta$ and $\phi$. The choice of $\theta$ and $\phi$ is dependent on $n_c$, which is equivalent to the network with $n_q+n_c$ nodes.

The experimental setups to simulate networks with $n_q+n_c$ nodes are shown in Fig.~\ref{fig:setup2}. For network with $n_q=1$, its equivalent quantum state is a single-photon state. As shown in Fig.~\ref{fig:setup2}a, triggering one photon of $\ket{GHZ}_2$ on $\ket{H}$ leaves the other photon on state $\ket{H}$. Then, by applying a QWP and HWP on it, arbitrary $\ket{\xi}_1$ can be generated. 

$\ket{\xi}_2$ is generated by overlapping one photon on $\ket{\xi}_1$ and another photon on $\ket{+}$ on a PBS (shown in Fig.~\ref{fig:setup2}b). Similar to the generation of $\ket{GHZ}_N$, $\ket{\xi}_{n_q}$ with $n_q=3, 4, 5$ can be generated with setups shown in Fig.~\ref{fig:setup2}c, d and e, respectively. 

\section{statistical significance}
In the main text, we analyze the standard deviation $\mathcal E$ of $F(N)$ in verifying entanglement and evaluating $n_c$, which is also related to statistical significance $\mathcal S$ \cite{Jungnitsch10}. The significance of $F(N)$ in EW is defined as $\mathcal S(EW)=(F(N)-0.5)/\mathcal E$, where 0.5 is the threshold of EW and $\mathcal E$ is the statistical error of $F(N)$ in our experiment. Similarly, the significance of $F(N)$ in evaluating $n_c$ is defined as $\mathcal S(n_c)=(F(N)-\mathcal F_{n_c-1})/\mathcal E$. The larger the statistical significance is, the higher confidence interval we can obtain regarding the conclusion.  The significance of $F(N)$ in networks with $n_q=1$ and $n_q=2$ are shown in Fig.~\ref{fig:data3}a and c, the value of significance of $F(N)$ in EW is greater than 3 and with maximal value of 68.5 when $n_c\leq6$, which represents a high confidence interval in EW. Again, it indicates EW is not confidential in identification of quantum network. With our criteria (shown in Figs. \ref{fig:data3}b and d), $F(N)$ can be employed as an confidential indicator in evaluating $n_c$ when $n_c\leq6$. Although $F(N)$ could exhibit graduation when $n_c\geq7$ (insets in Figs.~\ref{fig:data3}b and d), it can not be employed as a confidential indicator to evaluating $n_c$ as $\mathcal S(n_c)$ goes to subtle values.

\end{document}